\documentclass[preprint,1p,number]{elsarticle}

\usepackage{multirow}
\usepackage{graphicx}
\usepackage{dcolumn}
\usepackage{bm}
\usepackage{url}                                       
\usepackage{subfigure}
\usepackage{epstopdf}

\journal{Physica A}

\begin{document}

\begin{frontmatter}

\title{Sampling promotes community structure in social and information networks}
\author{Neli Blagus\corref{coraut}}
\ead{neli.blagus@fri.uni-lj.si}
\author{Lovro \v Subelj}
\ead{lovro.subelj@fri.uni-lj.si}
\author{Gregor Weiss}
\ead{gregor.weiss@fri.uni-lj.si}
\author{Marko Bajec}
\ead{marko.bajec@fri.uni-lj.si}
\address{University of Ljubljana, Faculty of Computer and Information Science, Ljubljana, Slovenia}
\cortext[coraut]{Corresponding author. Tel.: +386 1 476 81 86.}

\begin{abstract}
Any network studied in the literature is inevitably just a sampled representative of its real-world analogue. Additionally, network sampling is lately often applied to large networks to allow for their faster and more efficient analysis. Nevertheless, the changes in network structure introduced by sampling are still far from understood. In this paper, we study the presence of characteristic groups of nodes in sampled social and information networks. We consider different network sampling techniques including random node and link selection, network exploration and expansion. We first observe that the structure of social networks reveals densely linked groups like communities, while the structure of information networks is better described by modules of structurally equivalent nodes. However, despite these notable differences, the structure of sampled networks exhibits stronger characterization by community-like groups than the original networks, irrespective of their type and consistently across various sampling techniques. Hence, rich community structure commonly observed in social and information networks is to some extent merely an artifact of~sampling.
\end{abstract}


\end{frontmatter}

\section{\label{sec:intro}Introduction}
Any network found in the literature is inevitably just a sampled representative of its real-world analogue under study. For instance, many networks change quickly over time and in most cases merely incomplete data is available on the underlying system. Additionally, network sampling techniques are lately often applied to large networks to allow for their faster and more efficient analysis. Since the findings of the analyses and simulations on such sampled networks are implied for the original ones, it is of key importance to understand the structural differences between the original networks and their sampled variants.

A large number of studies on network sampling focused on the changes in network properties introduced by sampling. Lee~et~al.~\cite{LKJ06} showed that random node and link selection overestimate the scale-free exponent~\cite{BA99} of the degree and betweenness centrality~\cite{Fre77} distributions, while they preserve the degree mixing~\cite{Newman02}. On the other hand, random node selection preserves the degree distribution of different random graphs~\cite{SWM05} and performs better for larger sampled networks~\cite{SCBFGP12}. Furthermore, Leskovec~et~al.~\cite{LF06} showed that the exploration sampling using random walks or forest-fire strategy~\cite{LKF05} outperforms the random selection techniques in preserving the clustering coefficient~\cite{WS98}, different spectral properties~\cite{LF06}, and the in-degree and out-degree distributions. More recently, Ahmed~et~al.~\cite{ANK11} proposed random link selection with additional induction step, which notably improves on the current state-of-the-art. Their results confirm that the proposed technique well captures the degree distributions, shortest paths~\cite{WS98} and also the clustering coefficient of the original networks. Lately, different studies also focus on finding and correcting biases in sampling process, for example observing the changes of user attributes under the sampling of social networks~\cite{PM13}, analyzing the bias of traceroute sampling~\cite{LBCX03} and understanding the changes of degree distribution and hubs inclusion under various sampling techniques~\cite{MB11}. However, despite all those efforts, the changes in network structure introduced by sampling and the effects of network structure on the performance of sampling are still far from understood.

Real-world networks commonly reveal communities (also link-density community~\cite{LWZY07}), described as densely connected clusters of nodes that are loosely connected between~\cite{GN02}. Communities possibly play important roles in different real-world systems, for example in social networks communities represent friendship circles or people with similar interest~\cite{SC11}, while in citation networks communities can help us to reveal relationships between scientific disciplines~\cite{RB08}. Furthermore, community structure has a strong impact on dynamic processes taking place on networks~\cite{ADP06} and thus provides an important insight into structural organization and functional behavior of real-world systems. Consequently, a number of community detection algorithms have been proposed over the last years~\cite{WH04,RB07,RAK07,SB11} (for a review see~\cite{Fortunato10}). Most of these studies focus on classical communities characterized by higher density of edges~\cite{RCCLP04}. However, some recent works demonstrate that real-world networks reveal also other characteristic groups of nodes~\cite{NL07,PSR10} like groups of structurally equivalent nodes denoted modu\-les~\cite{NL07,SB12} (also link-pattern community~\cite{LWZY07} and other~\cite{RW07}), or different mixtures of communities and modules~\cite{SBB13}. 

Despite community structure appears to be an intrinsic property of many real-world networks, only a few studies considered the effects between the community structure and network sampling. Salehi et al.~\cite{SRR12} proposed Page-Rank sampling, which improves the performance of sampling of networks with strong community structure. Furthermore, expansion sampling~\cite{MBT10} directly constructs a sample representative of the community structure, while it can also be used to infer communities of the unsampled nodes. Other studies, for example analyzed the evolution of community structure in collaboration networks and showed that the number of communities and their size increase over time~\cite{LLPP14}, while the network sampling has a potential application in testing for signs of preferential attachment in the growth of networks~\cite{perc14}. However, to the best of our knowledge, the question whether sampling destroys the structure of communities and other groups of nodes or are sampled nodes organized in a similar way than nodes in original network remains unanswered.

In this paper, we study the presence of characteristic groups of nodes in different social and information networks and analyze the changes in network group structure introduced by sampling. We consider six sampling techniques including random node and link selection, network exploration and expansion sampling. The results first reveal that nodes in social networks form densely linked community-like groups, while the structure of information networks is better described by modules. However, regardless of the type of the network and consistently across different sampling techniques, the structure of sampled networks exhibits much stronger characterization by community-like groups than the original networks. We therefore conclude that the rich community structure is not necessary a result of for example homophily in social networks.

The rest of the paper is structured as follows. In Section~\ref{sec:sampl}, we introduce different sampling techniques considered in the study, while the adopted node group extraction framework is presented in Section~\ref{sec:nodegroups}. The results of the empirical analysis are reported and formally discussed in Section~\ref{sec:analys}, while Section~\ref{sec:conclusion} summarizes the paper and gives some prominent directions for future research.

\section{\label{sec:sampl}Network sampling}
\begin{figure}[!t]
\centering
\subfigure[RND]{\includegraphics[width=0.3\columnwidth]{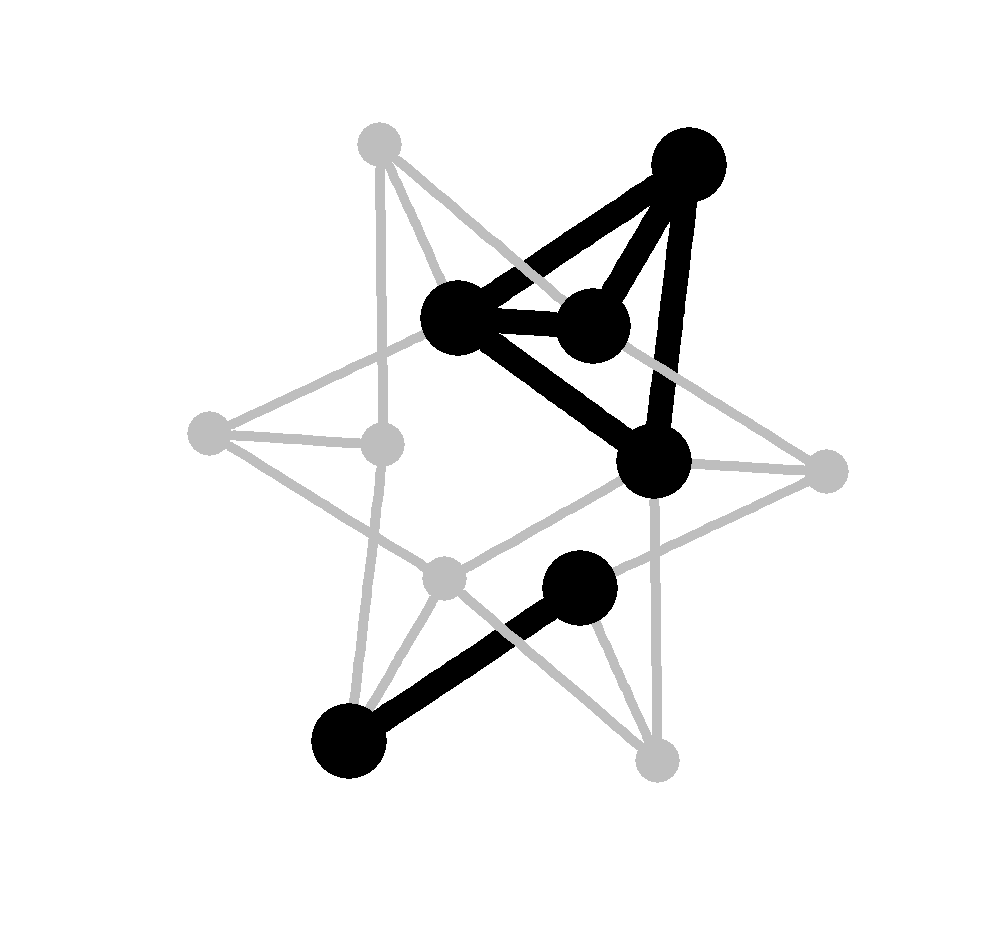}\label{subfig:RND}} \quad
\subfigure[RLS]{\includegraphics[width=0.3\columnwidth]{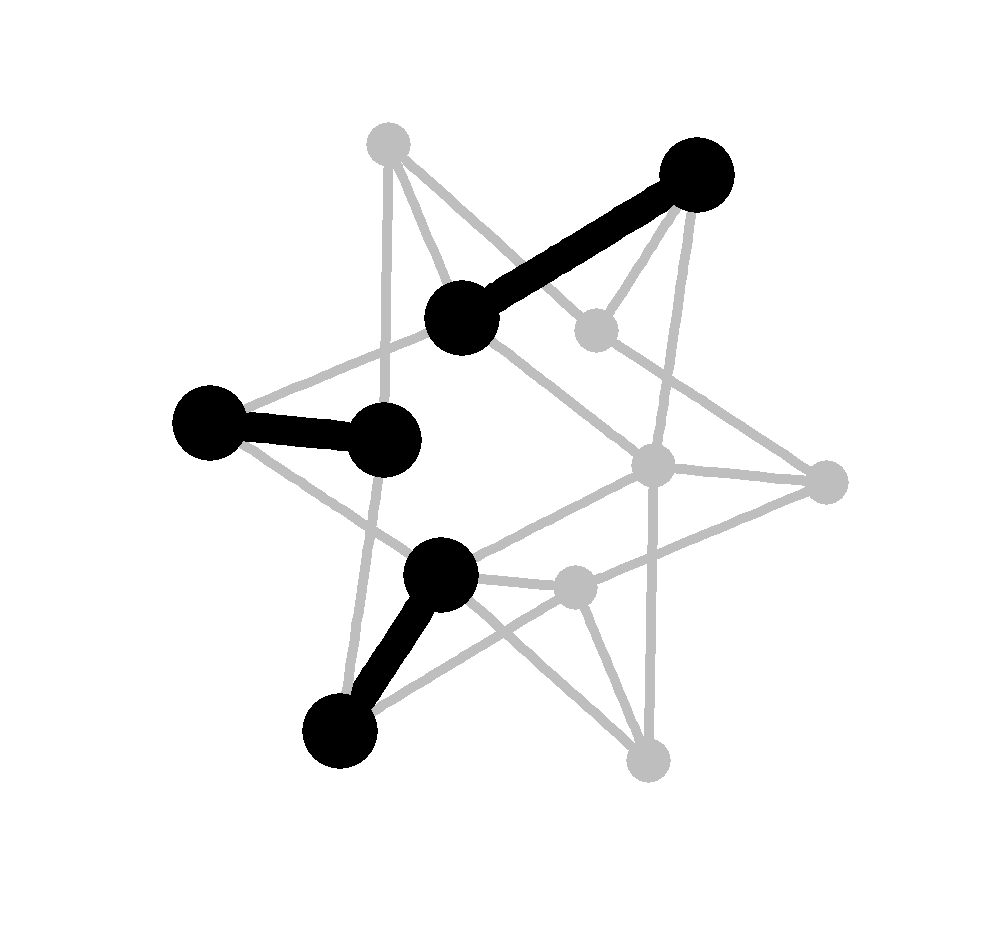}\label{subfig:RLS}} \quad
\subfigure[RLI]{\includegraphics[width=0.3\columnwidth]{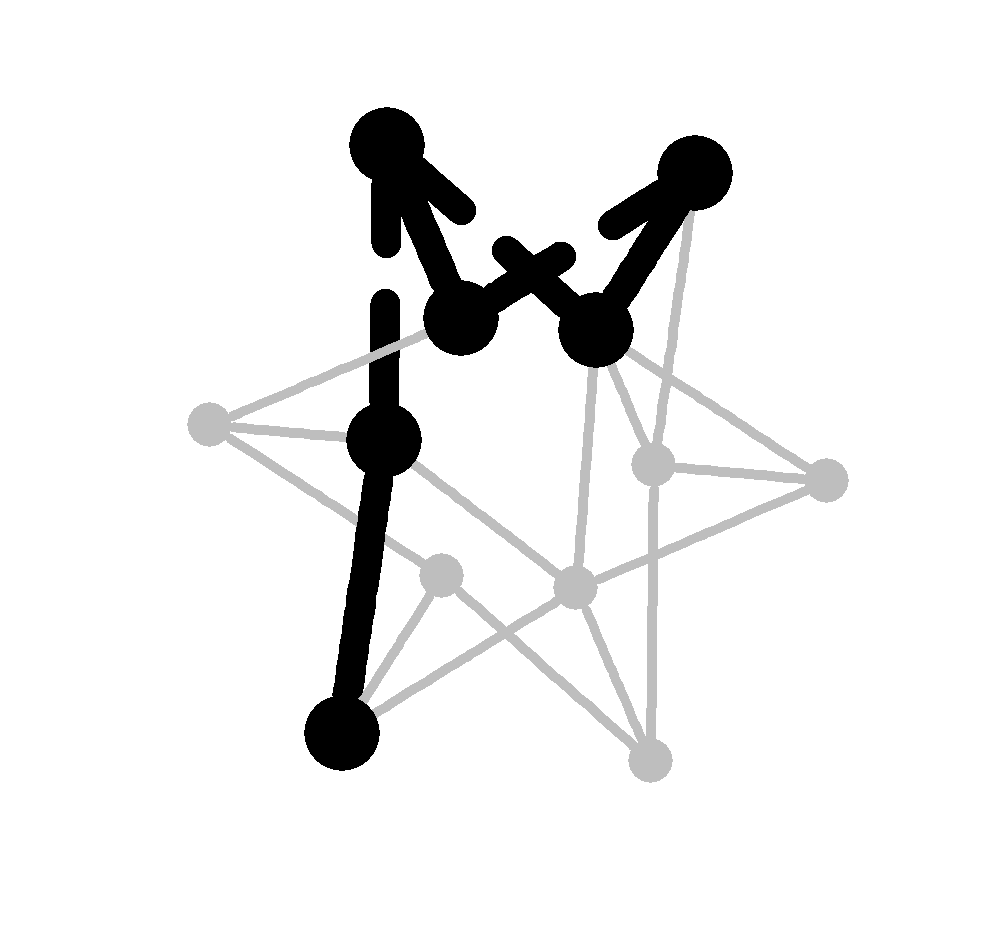}\label{subfig:RLI}}
\caption{Random selection techniques applied to a small toy network, where the samples are shown with highlighted nodes and links. \subref{subfig:RND}~In random node selection by degree, the nodes are selected with probability proportional to their degrees, while their mutual links are included in the sample. \subref{subfig:RLS}~In random link selection, the sample consists of links selected uniformly at random. \subref{subfig:RLI}~In random link selection with induction, the sample consists of randomly selected links (solid lines) and the links between their endpoints (dashed lines).}
\label{fig:selection}
\end{figure}

Network sampling techniques can be roughly divided into two categories: random selection and network exploration techniques. In the first category, nodes or links are included in the sample uniformly at random or proportional to some particular characteristic like the degree of a node or its PageRank score~\cite{BP98}. In the second category, the sample is constructed by retrieving a neighborhood of a randomly selected seed node using random walks, breadth-first search or another strategy. For the purpose of this study, we consider three techniques from each of the~categories.

\subsection{Random selection}
From the random selection category, we first adopt random node selection by degree~\cite{LF06} (RND). Here, the nodes are selected randomly with probability proportional to their degrees, while all their mutual links are included in the sample (Fig.~\ref{subfig:RND}). Note that RND improves the performance of the basic random node selection~\cite{LF06,BSB14}, where the nodes are selected to the sample uniformly at random. RND fits better spectral network properties~\cite{LF06} and produces the sample with larger weakly connected component~\cite{BSB14}. Moreover, it shows good performance in preserving the clustering coefficient and betweenness centrality distribution of the original networks~\cite{BSB14}. Nevertheless, it can still construct a disconnected sample network, despite a fully connected original network.

Next, we adopt random link selection~\cite{LF06} (RLS), where the sample consists of links selected uniformly at random (Fig.~\ref{subfig:RLS}). RLS overestimates degree and betweenness centrality exponent, underestimate the clustering coefficient and accurately matches the assortativity of the original network~\cite{LKJ06}. The samples created with RLS are sparse and the connectivity of the original network is not preserved, still RLS is likely to capture the path length of the original network~\cite{ANK12}.

Last, we adopt random link selection with induction~\cite{ANK11} (RLI), which improves the performance of RLS. In RLI, the sample consists of randomly selected links as before, while also all additional links between their endpoints (Fig.~\ref{subfig:RLI}). RLI outperforms several other methods in capturing the degree, path length and clustering coefficient distribution. It selects nodes with higher degree than RLS, thus the connectivity of the sample is increased~\cite{ANK11}.

Techniques from random selection category imitate classical statistical sampling approaches, where each individual is selected from population independently from others until desired size of the sample is reached. 

\subsection{Network exploration}
\begin{figure}[t]
\centering
\subfigure[BFS]{\includegraphics[width=0.3\columnwidth]{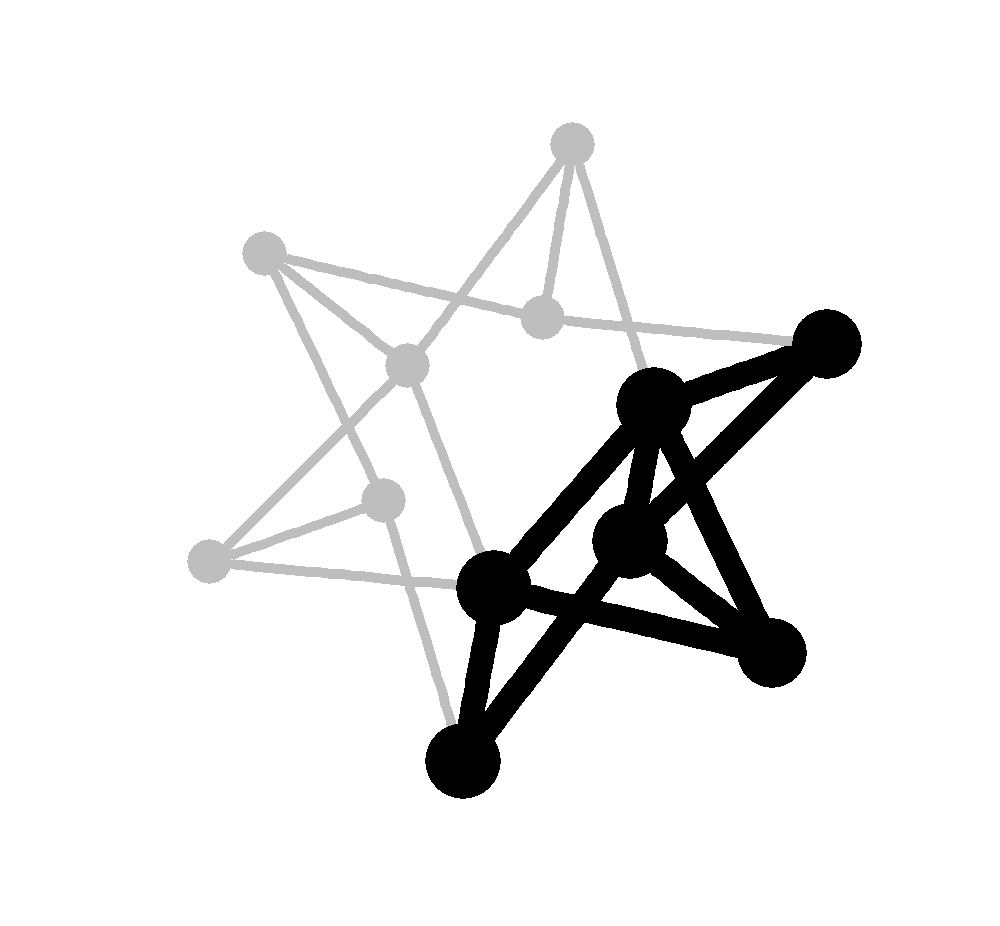}\label{subfig:BFS}} \quad
\subfigure[FFS]{\includegraphics[width=0.3\columnwidth]{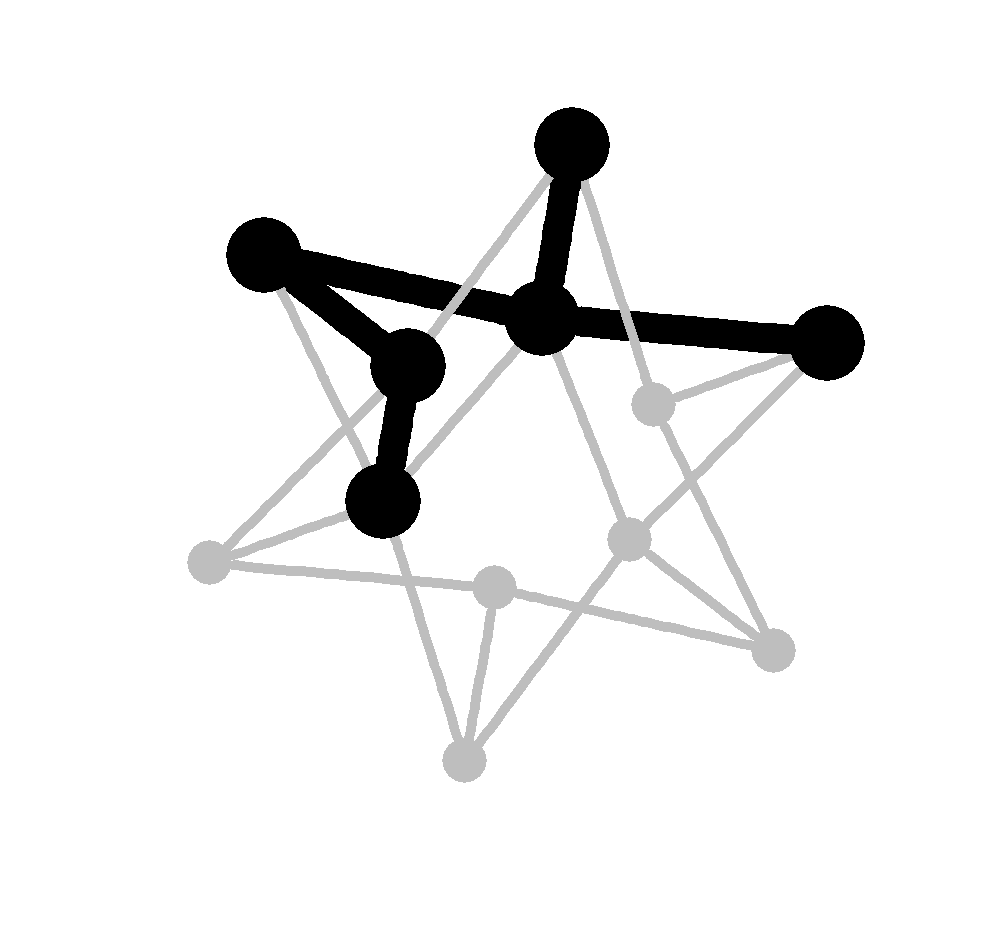}\label{subfig:FFS}} \quad
\subfigure[EXS]{\includegraphics[width=0.3\columnwidth]{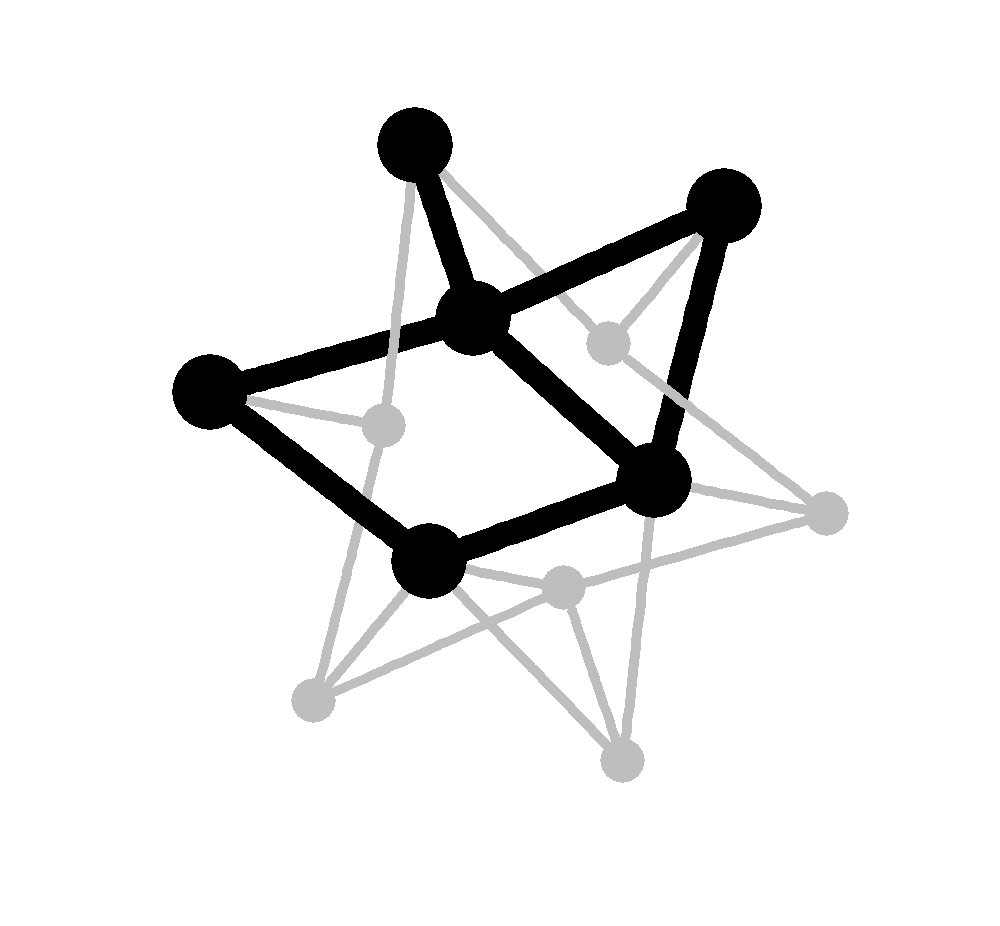}\label{subfig:EXS}}
\caption{Network exploration techniques applied to a small toy network, where the samples are shown with highlighted nodes and links. \subref{subfig:BFS}~In breadth-first sampling, a seed node is first selected uniformly at random, while its broad neighborhood retrieved from breadth-first search is included in the sample. \subref{subfig:FFS}~In forest-fire sampling, the broad neighborhood of a randomly selected seed node is retrieved from partial breadth-first search, where only a fraction of neighbors is included in the sample. \subref{subfig:EXS}~In expansion sampling, the seed node is selected uniformly at random, while the remaining nodes are selected from the neighborhood of sampled nodes with probability proportional to their contribution to the expansion factor (see text).}
\label{fig:exploration}
\end{figure}

From the network exploration category, we first adopt breadth-first sampling~\cite{LKJ06} (BFS). Here, a seed node is selected uniformly at random, while its broad neighborhood retrieved from the basic breadth-first search is included in the sample (Fig.~\ref{subfig:BFS}). The sample network is thus a connected subgraph of the original network. BFS is biased towards selecting high-degree nodes in the sample~\cite{KMT10}. It captures well the degree distribution of the networks, while it performs worst in inclusion of hubs in the sample quickly in the sampling process~\cite{MB11}. BFS imitates the snowball sampling approach for collecting social data used especially when the data is difficult to reach~\cite{goodman61}. Selected seed participant is asked to report his friends, which are than invited to report their friends. The procedure is repeated until the desired number of people is sampled.

Next, we adopt a modification of BFS denoted forest-fire sampling~\cite{LF06} (FFS). In FFS, the broad neighborhood of a randomly selected seed node is retrieved from partial breadth-first search, where only some neighbors are included in the sample on each step (Fig.~\ref{subfig:FFS}). The number of neighbors is sampled from a geometric distribution with mean $p/(1-p)$, where $p$ is set to $0.7$~\cite{LF06}. FFS matches well spectral properties~\cite{LF06}, while it underestimates the degree distribution and fails to match the path length and clustering coefficient of the original networks~\cite{ANK12}. However, FFS corresponds to a model by which one author collects the papers to cite and include them in the bibliography~\cite{LKF05}. The author starts with one paper, explores its bibliography and selects the papers to cite. The procedure is recursively repeated in selected papers until desired collection of citations is reached.

Last, we adopt expansion sampling~\cite{MBT10} (EXS), where the seed node is again selected uniformly at random, while the neighbors of the sampled nodes are included in the sample with probability proportional to
\begin{equation}
1-\beta^{|N(\{v\})-(N(S) \cup S)|},
\end{equation}
where $v$ is the concerned node, $S$ the current sample and $N(S)$ the neighborhood of nodes in $S$ (Fig.~\ref{subfig:EXS}). Expression $|N(\{v\})-(N(S) \cup S)|$ denotes the expansion factor of node $v$ for sample $S$ and means the number of new neighbors contributed by $v$. The parameter $\beta$ is set to $0.9$~\cite{MBT10}. Note that EXS ensures that the sample consists of nodes from most communities in the original network and that the nodes that are grouped together in the original network, are also grouped together in the sample~\cite{GN02}. EXS imitates the modification of snowball sampling approach mentioned above, where for example we want to gather the data about individuals from different countries. Thus, on each step we include in the sample the individuals, which knows larger number of others from various countries.

\section{\label{sec:nodegroups}Group extraction}
The node group structure of different networks is explored by a group extraction framework~\cite{SBB13,ZLZ11,Weiss} with a brief overview below.

Let the network be represented by an undirected graph $G(V, L)$, where $V$ is the set of nodes and $L$ the set of links. Next, let $S$ be a group of nodes and $T$ a subset of nodes representing its corresponding linking pattern (i.e., the pattern of connections of nodes from $S$ to other nodes~\cite{NL07}), $S,T\subseteq V$. Denote $s=|S|$ and $t=|T|$. The linking pattern $T$ is selected to maximize the number of links between $S$ and $T$, and minimize the number of links between $S$ and $T^C$, while disregarding the links with both endpoints in $S^C$. For details on the group objective function see~\cite{SBB13,SZBB14}.

\begin{figure}[!t]
\centering
\subfigure[Community]{\includegraphics[width=0.3\columnwidth]{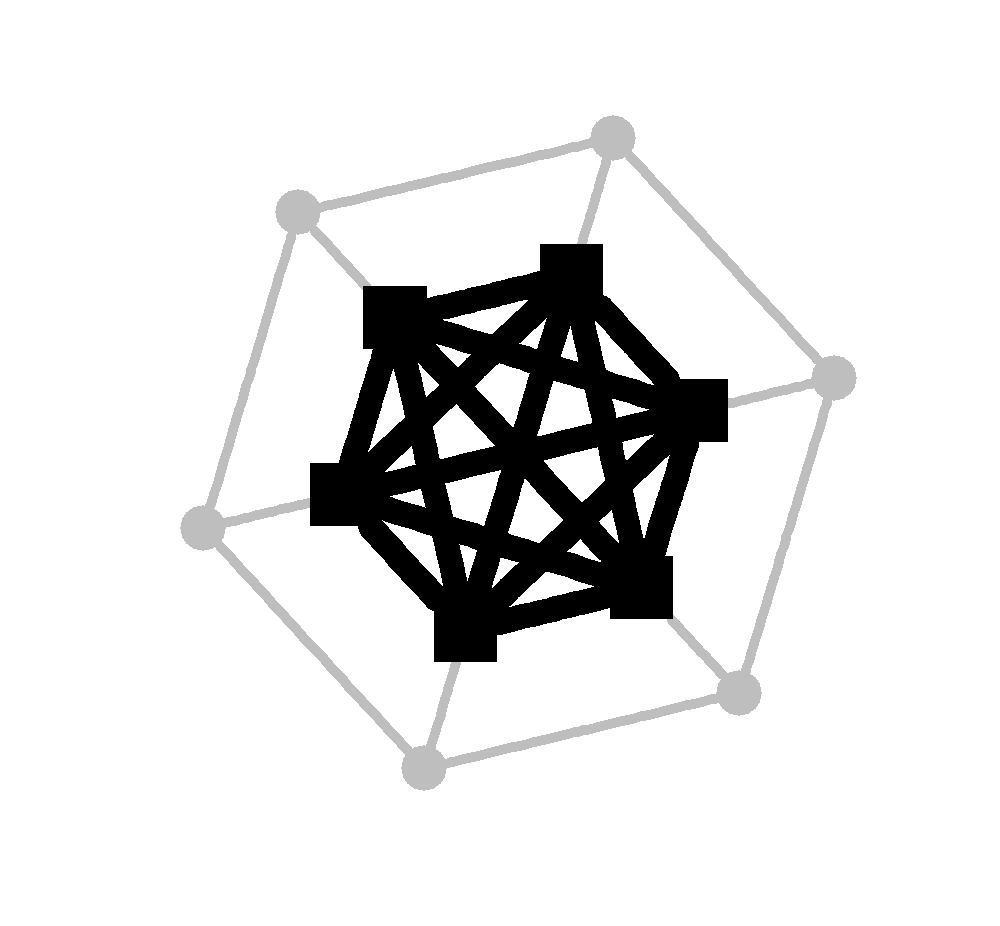}\label{subfig:comm}}\quad
\subfigure[Module]{\includegraphics[width=0.3\columnwidth]{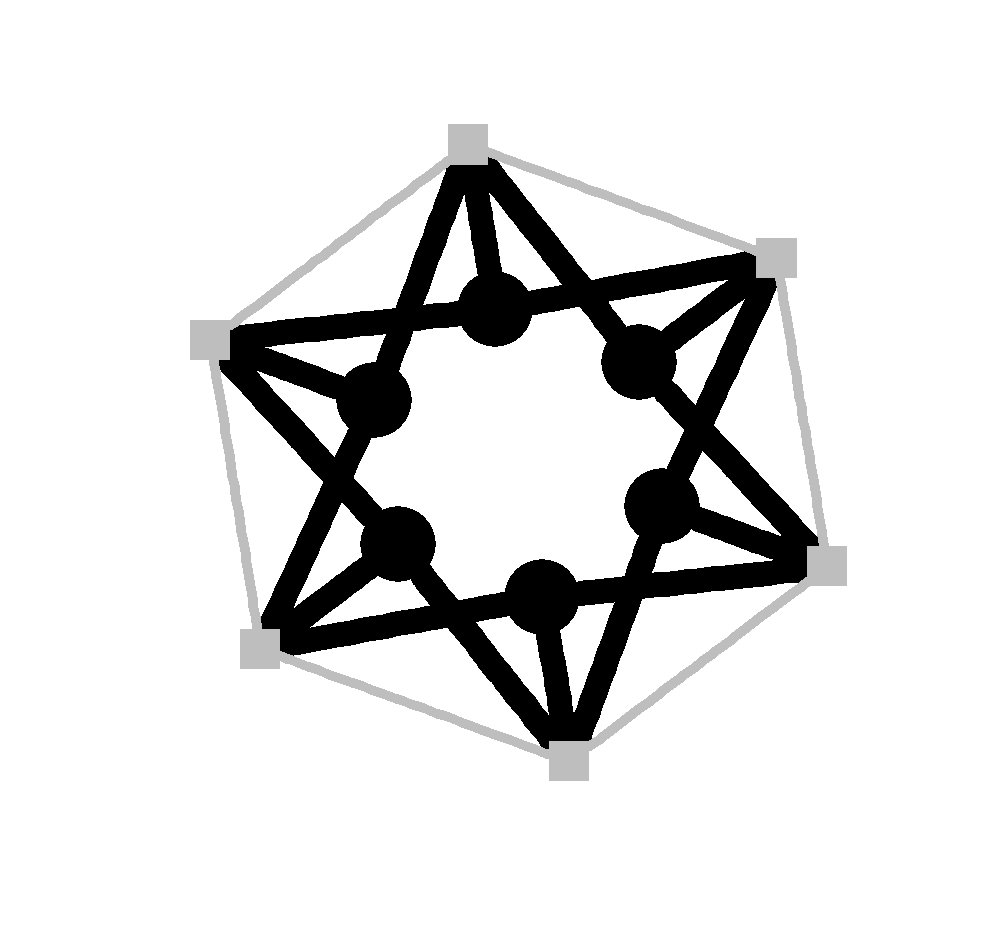}\label{subfig:mod}}
\caption{Toy examples of groups of nodes in networks, where groups $S$ and their corresponding linking patterns $T$ are shown with highlighted and squared nodes, respectively (see text). \subref{subfig:comm}~Communities are densely connected groups of nodes with $S=T$. \subref{subfig:mod}~Modules are possibly disconnected groups of structurally equivalent nodes with $S\cap T=\emptyset$. Groups spanning between communities and modules are denoted mixtures.}
\label{fig:groups}
\end{figure}

The above formalism comprises different types of groups commonly analyzed in the literature (Fig.~\ref{fig:groups}). It consider communities~\cite{GN02} (i.e., link-density community~\cite{LWZY07}), defined as a (connected) group of nodes with more links toward the nodes in the group than to the rest of the network~\cite{RCCLP04}. Communities are characterized by $S=T$. Furthermore, the formalism consider possibly disconnected groups of structurally equivalent nodes denoted modules~\cite{NL07,SB12} (i.e., link-pattern community~\cite{LWZY07}), defined as a (possibly) disconnected group of nodes with more links towards common neighbors than to the rest of the network~\cite{RCCLP04}. Modules have $S \cap T = \emptyset$. Communities and modules represent two extreme cases with all other groups being the mixtures of the two~\cite{SBB13}, $S \cap T \subset S$ and/or $S \cap T \subset T$. The reader may also find it interesting that the core-periphery structure is a mixture with $S\subset T$, while the hub \& spokes structure is a module with $t=1$.

The type of group $S$ can in fact be determined by the Jaccard index~\cite{jaccard1901} of $S$ and its corresponding linking pattern $T$. The group parameter $\tau$~\cite{SBB13}, $\tau \in [0,1]$, is defined as
\begin{equation}
\tau(S,T)=\frac{|S \cap T|}{|S \cup T|}.
\end{equation}
Communities have $\tau = 1$, while modules are indicated by $\tau = 0$. Mixtures correspond to groups with $0 <\tau < 1$. For the rest of the paper, we refer to groups with $\tau\approx 1$ as community-like and groups with $\tau\approx 0$ as module-like.

Groups in networks are revealed by a sequential extraction procedure proposed in~\cite{ZLZ11,SBB13,Weiss}. One first finds the group $S$ and its linking pattern $T$ with random-restart hill climbing~\cite{RN03} that maximizes the objective function. Next, the revealed group $S$ is extracted from the network by removing the links between groups $S$ and $T$, and any node that becomes isolated. The procedure is then repeated on the remaining network until the objective function is larger than the $99$th percentile of the values obtained under the same framework in a corresponding Erd\H{o}s-R{\'e}nyi random graph~\cite{ER59}. All groups reported in the paper are thus statistically significant at $1\%$~level. Note that the above procedure allows for overlapping~\cite{PDFV05}, hierarchical~\cite{RSMOB02}, nested and other classes of groups.

\section{\label{sec:analys}Analysis and discussion}
Section~\ref{subsec:nets} introduces real-world networks considered in the study. Section~\ref{subsec:orig} reports the node group structure of the original networks extracted with the framework described in Section~\ref{sec:nodegroups}. The groups extracted from the sampled networks are analyzed in Section~\ref{subsec:sampled}. For a complete analysis, we also observe the node group structure of a large network with more than a million links in Section~\ref{subsec:large}.

\subsection{\label{subsec:nets}Network data}

\begin{table}[!t]
\scriptsize
\centering
\caption{Social and information networks considered in the study.}
\label{tbl:nets}
\begin{tabular}{clrr}
\hline\noalign{\smallskip}
\multicolumn{1}{c}{Network} & \multicolumn{1}{c}{Description} &  \multicolumn{1}{c}{\# Nodes} &  \multicolumn{1}{c}{\# Links} \\
\noalign{\smallskip}\hline\noalign{\smallskip}
\textit{Collab} & High Energy Physics collaborations~\cite{LKF05} & $9877$ & $25998$ \\
\textit{PGP} & Pretty Good Privacy web-of-trust~\cite{BPDA04} & $10680$ & $24340$ \\
\noalign{\smallskip}\hline\noalign{\smallskip}
\textit{P2P} & Gnutella peer-to-peer file sharing~\cite{LKF05} & $8717$ & $31525$ \\
\textit{Citation} &  High Energy Physics citations~\cite{LKF05} & $27770$ & $352807$ \\
\noalign{\smallskip}\hline
\end{tabular}
\end{table}

The empirical analysis in the following sections was performed on four real-world social and information networks. Their main characteristics are shown in Table~\ref{tbl:nets}.

The \textit{Collab}~\cite{LKF05} is a social network of scientific collaborations among researchers, who submitted their papers to High Energy Physics -- Theory category on the arXiv in the period from January $1993$ to April $2003$. The nodes represent the authors, while undirected links denote that two authors co-authored at least one paper together.

The \textit{PGP}~\cite{BPDA04} is a social network, which corresponds to the interaction network of users of the Pretty Good Privacy algorithm collected in July $2001$. The nodes represent users, while undirected links indicate relationships between those, who sign each other's public key.

The \textit{P2P}~\cite{LKF05} is an information network, which contains a sequence of snapshots of the Gnu\-te\-lla peer-to-peer file sharing network collected in August $2002$. The nodes represent hosts in the Gnutella network, which are linked by undirected links if there exist connections between them.

The \textit{Citation}~\cite{LKF05} is an information network, again gathered from the High Energy Physics -- Theory category from the arXiv in the period from January $1993$ to April $2003$ and includes the citations among all papers in the dataset. The network consists of nodes, which represent papers, while links denote that one paper cite another.

\subsection{\label{subsec:orig}Group structure of original networks}
We first analyze the properties of groups extracted from the original networks summarized in Table~\ref{tbl:orig}.

The number of groups differs among networks, still the mean group size $s$ (denoted $\left<s\right>$) is comparable across network types. Groups $S$ in social networks consist of around $64$ nodes, while $\left<s\right>$ in information networks exceeds $150$ nodes. The mean linking pattern size $t$ (denoted $\left<t\right>$) of social networks is comparable to $\left<s\right>$. The latter relation $\left<t\right> \approx \left<s\right>$ is expected due to the pronounced community structure commonly found in social networks~\cite{NP03}. On the other hand, $\left<t\right> > \left<s\right>$ is expected for information networks, due to the abundance of module-like groups.

The characteristic group structure of networks is reflected in the group parameter $\tau$. For social networks, its values are around $0.556$, which indicates the presence of communities, modules and mixtures of these. In contrast to social networks, the information networks have $\tau$ closer to $0$ and consist mostly of module-like groups.

To summarize, social networks represent people and interactions between them, like a few authors writing a paper together, therefore we can expect a larger number of community-like groups in these networks. On the other hand, in information networks the homophily is less typical and thus the structure of these networks seem better described by module-like groups.

\begin{table}[!t]
\scriptsize
\centering
\caption{Groups of nodes extracted from social and information networks. We report the number of groups \#, the mean group size $s$, the mean pattern size $t$, the mean group parameter $\tau$, the median group parameter denoted $m_{\tau}$ and the distribution over different types of groups (see text). Notice that social networks consist of smaller groups with larger $\tau$ than information~networks.
}
\label{tbl:orig}
\begin{tabular}{crrrrrrrr}
\hline\noalign{\smallskip}
\multicolumn{1}{c}{Network} & \multicolumn{5}{c}{Group} & \multicolumn{1}{c}{Community} & \multicolumn{1}{c}{Mixture} & \multicolumn{1}{c}{Module}\\
& \multicolumn{1}{c}{\#} & \multicolumn{1}{c}{$\left\langle s \right\rangle$} & \multicolumn{1}{c}{$\left\langle t \right\rangle$} & \multicolumn{1}{c}{$\left\langle \tau \right\rangle$} & \multicolumn{1}{c}{$m_{\tau}$}& \multicolumn{3}{c}{Distribution \%} \\
\noalign{\smallskip}\hline\noalign{\smallskip}
\textit{Collab} & $129$ & $66.9$  & $67.2$ & $0.568$ & $0.554$ & $1.6 \%$ & $96.8\%$ & $1.6\%$ \\
\textit{PGP} & $87$ & $62.2$ & $61.9$ & $0.568$ & $0.536$ & $4.6 \%$ & $94.3\%$ & $1.1\%$ \\
\noalign{\smallskip}\hline\noalign{\smallskip}
\textit{P2P} & $70$ & $154.8$  & $177.0$ & $0.057$ & $0.000$ & $0.0 \%$ & $44.3 \%$ & $55.7 \%$ \\
\textit{Citation} & $284$ & $271.7$  & $280.6$ & $0.186$ & $0.120$ & $0.0 \%$ & $96.8 \%$ & $3.2 \%$ \\
\noalign{\smallskip}\hline
\end{tabular}
\end{table}

\subsection{\label{subsec:sampled}Group structure of sampled networks}
Sampling techniques outlined in Section~\ref{sec:sampl} enable setting the size of the sampled networks in advance. We consider sample sizes of $15\%$ of nodes from the original networks, that provides for an accurate fit of several network properties~\cite{LF06,BSB14}. 

Table~\ref{tbl:samplsoc} and~\ref{tbl:samplinf} present the properties of the node group structure of sampled social and information networks, respectively. Notice that RLS and FFS show different performance than other techniques. The samples obtained with RLS and FFS contain less groups, which consist of no more than $36$ nodes. Additionally, almost all groups in these samples are modules, which reflects in the mean group parameter $\tau$ (denoted $\left<\tau\right>$) approaching $0$ for all networks.

To verify the above findings, we compute externally studentized residuals of the sampled networks that measure the consistency of each sampling technique with the rest. The residuals are calculated for each technique as the difference between the observed value of considered property and its mean divided by the standard deviation. The mean value and standard deviation are computed for all sampling techniques, excluding the observed one (for details see~\cite{SFB14}). Statistically significant inconsistencies between techniques are revealed by two-tailed Student $t-$test~\cite{CW82} at $P-$value of $0.1$, rejecting the null hypothesis that the values of the considered property are consistent across the sampling techniques.

\begin{figure}[!t]
\centering
\subfigure[Number of groups \#]{\includegraphics[width=0.45\columnwidth]{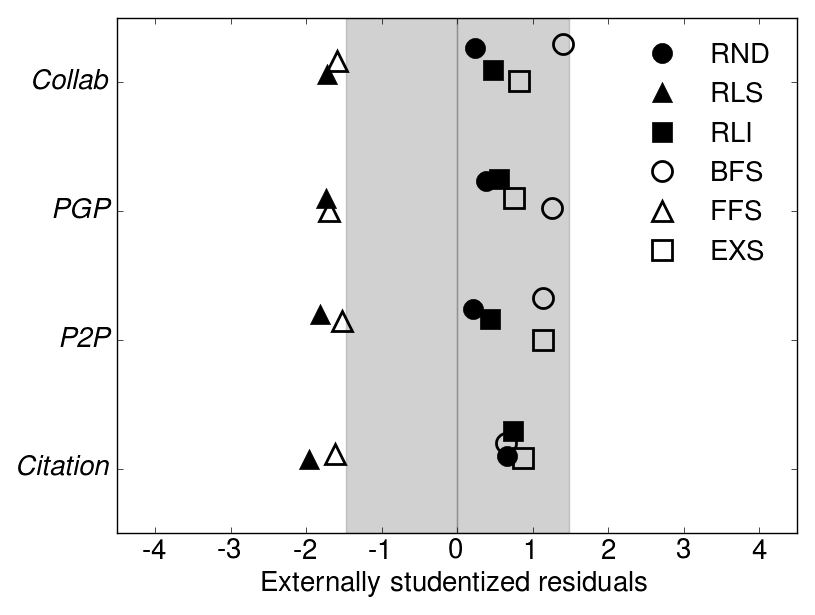}\label{subfig:nr}} \quad
\subfigure[Group parameter $\left< \tau \right>$]{\includegraphics[width=0.45\columnwidth]{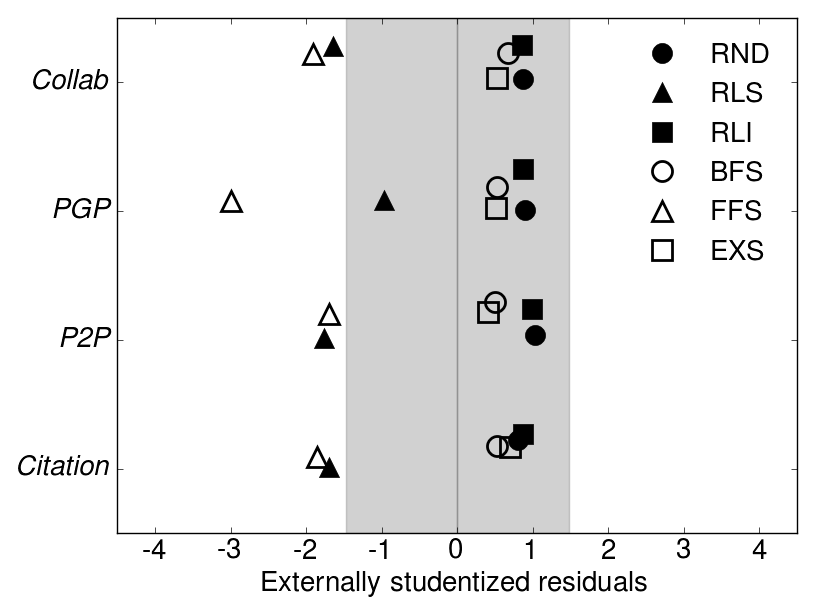}\label{subfig:tau}} \\
\caption{Statistical comparison of \subref{subfig:nr} number of groups and \subref{subfig:tau} mean group parameter $\tau$ for the sampled networks obtained with different sampling techniques (see text). We show externally studentized residuals that measure the consistency of each sampling technique with the rest and expose statistically significant inconsistencies between the techniques with two-tailed Student $t$-test at $P\mbox{-value}$ of $0.1$ (shaded regions correspond to $90\%$ confidence intervals). Notice that sampled networks obtained with RLS and FFS reveal less groups (see \subref{subfig:nr}) with significantly smaller parameter $\tau$ (see \subref{subfig:tau}) than other sampling techniques.
}
\label{fig:stau}
\end{figure}

Statistical comparison of sampling techniques for the number of groups and the mean group parameter $\tau$ is shown on Fig.~\ref{fig:stau}. We confirm that the samples obtained with RLS and FFS reveal significantly less groups with significantly smaller $\left<\tau\right>$ than other sampling techniques. Moreover, if we compare the number of links in the sampled networks, RLS and FFS create samples that contain on average $3\%$ of links from the original networks. In contrast, the samples obtained with RND, RLI, BFS and EXS consist of around $16\%$ of links from the original networks. As mentioned before, the sizes of all samples are $15\%$ of the original networks, thus the sampled networks obtained with RLS and FFS are much sparser than others. In addition, the performance of RLS and FFS can also be explained by their definition. Since in RLS we include only randomly selected links in the sample, the variance is very high, while it commonly contains a large number of sparsely linked components, whose structure is best described as module-like. On the other hand, the samples obtained with FFS consist of one connected component with a low average degree of $2.33$. Thus, the sparsely connected nodes also form groups, which are more similar to modules. Due to the above reasons, we exclude RLS and FFS from further analysis. We focus on RND, RLI, BFS, and EXS, whose performance is clearly more comparable.

The selected sampling techniques perform similarly ac\-ross all networks as shown in Table~\ref{tbl:samplsoc} for social and Table~\ref{tbl:samplinf} for information networks. The samples consist of various number of groups, still in most cases less than the original networks. The mean sizes $s$ and $t$ are around $40$, in contrast to groups with $143$ nodes on average in the original networks.  Still, $\left<s\right> \approx \left<t\right>$ irrespective of network type and the sampling technique, which implies stronger characterization by community-like groups, as already argued in the case of social networks in Section~\ref{subsec:orig}.

\begin{table}[!t]
\scriptsize
\centering
\caption{Groups of nodes extracted from sampled social networks over $100$ realizations of different sampling techniques (see text). We report the number of groups \# and standard deviation, the mean group size $s$, the mean pattern size $t$, the mean group parameter $\tau$ and standard deviation, the median group parameter denoted $m_{\tau}$ and the distribution over different types of groups. Notice that sampled networks expectedly consist of smaller groups, but with larger $\tau$ than original social~networks (see $\left\langle \tau \right\rangle$ and $m_{\tau}$).}
\label{tbl:samplsoc}
\begin{tabular}{ccrrrrrrrrrr}
\hline\noalign{\smallskip}
\multicolumn{1}{c}{Network} & \multicolumn{1}{c}{Sampling} & \multicolumn{5}{c}{Group} & \multicolumn{1}{c}{Community} & \multicolumn{1}{c}{Mixture} & \multicolumn{1}{c}{Module}\\
& & \multicolumn{1}{c}{\#} & \multicolumn{1}{c}{$\left\langle s \right\rangle$} & \multicolumn{1}{c}{$\left\langle t \right\rangle$} & \multicolumn{1}{c}{$\left\langle \tau \right\rangle$} &  \multicolumn{1}{c}{$m_{\tau}$} & \multicolumn{3}{c}{Distribution \%} \\
\noalign{\smallskip}\hline\noalign{\smallskip}
\multirow{7}{*}{\textit{Collab}} & / &  \multicolumn{1}{l}{$129.0$} & $66.9$  & $67.2$ &  \multicolumn{1}{l}{$0.568$} & $0.554$ & $1.6 \%$ & $96.8 \%$ & $1.6 \%$\\\noalign{\smallskip}\cline{2-10}\noalign{\smallskip}
& RND & $65.4\pm3.7$ & $13.5$  & $13.7$ & $0.851\pm0.030$& $0.989$ & $54.7 \%$ & $41.9 \%$ & $3.4 \%$  \\
& RLS & $1.2\pm0.5$ & $1.5$  & $4.8$ & $0.047\pm0.149$ & $0.048$ & $0.0 \%$ & $8.3 \%$ & $91.7 \%$ \\
& RLI & $74.7\pm4.6$ & $13.7$  & $13.9$ & $0.846\pm0.030$ & $0.979$ & $52.7 \%$ & $43.4 \%$ & $3.9 \%$ \\
& BFS & $104.0\pm6.5$ & $18.2$  & $18.5$ & $0.787\pm0.032$ & $0.861$ & $30.3 \%$ & $66.5 \%$ & $3.2 \%$ \\
& FFS & $4.0\pm1.6$ & $16.8$  & $29.8$ & $0.000\pm0.000$ & $0.000$ & $0.0 \%$ & $0.0 \%$ & $100.0 \%$  \\
& EXS & $87.0\pm5.8$ & $18.4$  & $18.9$ & $0.741\pm0.026$ & $0.791$ & $21.4 \%$  & $76.3 \%$ & $2.3 \%$ \\
\noalign{\smallskip}\hline\noalign{\smallskip}
\multirow{7}{*}{\textit{PGP}} & / &  \multicolumn{1}{l}{$\phantom{7}87.0$} & $62.2$ & $61.9$ &  \multicolumn{1}{l}{$0.568$} & $0.536$ & $4.6 \%$ & $94.3 \%$ & $1.1 \%$ \\\noalign{\smallskip}\cline{2-10}\noalign{\smallskip}
& RND & $68.2\pm4.5$ & $15.8$ & $16.0$ & $0.891\pm0.024$ & $1.000$ & $67.8 \%$ & $28.7 \%$ & $3.5 \%$ \\
& RLS & $2.8\pm1.0$ & $5.7$  & $7.6$ & $0.304\pm0.233$ & $0.263$ & $21.4 \%$ & $28.6 \%$ & $50.0 \%$ \\
& RLI & $74.3\pm4.3$ & $15.8$  & $16.1$ & $0.883\pm0.024$ & $1.000$ & $65.1 \%$ & $31.1 \%$ & $3.8 \%$ \\
& BFS & $95.4\pm9.2$ & $17.5$ & $17.7$ & $0.784\pm0.025$ & $0.909$ & $39.2 \%$ & $55.6 \%$ & $5.2 \%$ \\
& FFS & $3.6\pm1.3$ & $13.5$  & $32.6$ & $0.000\pm0.000$ & $0.000$ & $0.0 \%$ & $0.0 \%$ & $100.0 \%$ \\
& EXS & $80.9\pm6.5$ & $15.6$  & $15.8$ & $0.779\pm0.028$ & $0.873$ & $34.5 \%$ & $61.2 \%$ & $4.3 \%$ \\
\noalign{\smallskip}\hline
\end{tabular}
\end{table}

\begin{table}[!t]
\scriptsize
\centering
\caption{Groups of nodes extracted from sampled information networks over $100$ realizations of different sampling techniques (see text). We report the number of groups \# and standard deviation, the mean group size $s$, the mean pattern size $t$, the mean group parameter $\tau$ and standard deviation, the median group parameter denoted $m_{\tau}$ and the distribution over different types of groups. Notice that sampled networks expectedly consist of smaller groups, but with larger $\tau$ than original information~networks (see $\left\langle \tau \right\rangle$ and $m_{\tau}$).}
\label{tbl:samplinf}
\begin{tabular}{ccrrrrrrrrr}
\hline\noalign{\smallskip}
\multicolumn{1}{c}{Network} & \multicolumn{1}{c}{Sampling} & \multicolumn{5}{c}{Group} & \multicolumn{1}{c}{Community} & \multicolumn{1}{c}{Mixture} & \multicolumn{1}{c}{Module} \\
& & \multicolumn{1}{c}{\#} & \multicolumn{1}{c}{$\left\langle s \right\rangle$} & \multicolumn{1}{c}{$\left\langle t \right\rangle$} & \multicolumn{1}{c}{$\left\langle \tau \right\rangle$} &  \multicolumn{1}{c}{$m_{\tau}$} & \multicolumn{3}{c}{Distribution \%} \\
\noalign{\smallskip}\hline\noalign{\smallskip}
\multirow{7}{*}{\textit{P2P}} & / &  \multicolumn{1}{l}{$\phantom{7}70.0$} & $154.8$  & $177.0$ &  \multicolumn{1}{l}{$0.057$} & $0.000$ & $0.0 \%$ & $44.3 \%$ & $55.7 \%$ \\\noalign{\smallskip}\cline{2-10}\noalign{\smallskip}
& RND & $23.3\pm3.9$ & $24.2$  & $24.4$ & $0.163\pm0.049$ & $0.034$ & $4.2 \%$ & $45.8 \%$ & $50.0 \%$ \\
& RLS & $1.6\pm0.9$ & $1.2$  & $3.6$ & $0.000\pm0.008$ & $0.000$ & $0.0 \%$ & $0.0 \%$ & $100.0 \%$ \\
& RLI & $26.2\pm4.4$ & $27.5$  & $28.1$ & $0.161\pm0.039$ & $0.035$ & $3.8 \%$ & $48.8 \%$ & $47.4 \%$  \\
& BFS & $34.1\pm5.5$ & $31.3$  & $27.9$ & $0.131\pm0.042$ & $0.034$ & $2.3 \%$ & $50.7 \%$ & $47.0 \%$ \\
& FFS & $3.6\pm1.4$ & $17.8$  & $28.3$ & $0.000\pm0.000$ & $0.000$ & $0.0 \%$ & $0.0 \%$ & $100.0 \%$ \\
& EXS & $34.0\pm5.9$ & $36.9$  & $37.3$ & $0.125\pm0.030$ & $0.035$ & $2.4 \%$ & $53.8 \%$ & $43.8 \%$ \\
\noalign{\smallskip}\hline\noalign{\smallskip}
\multirow{7}{*}{\textit{Citation}} & / &  \multicolumn{1}{l}{$284.0$} & $271.7$  & $280.6$ &  \multicolumn{1}{l}{$0.186$} & $0.120$ & $0.0 \%$ & $96.8 \%$ & $3.2 \%$ \\\noalign{\smallskip}\cline{2-10}\noalign{\smallskip}
& RND & $121.4\pm4.9$ & $74.9$  & $78.1$ & $0.405\pm0.016$ & $0.329$ & $0.2 \%$ & $80.9 \%$ & $18.9 \%$ \\
& RLS & $1.5\pm1.2$ & $1.4$  & $15.3$ & $0.014\pm0.073$ & $0.014$ & $0.0 \%$ & $0.0 \%$ & $100.0 \%$ \\
& RLI & $124.8\pm5.5$ & $76.3$  & $79.9$ & $0.415\pm0.014$ & $0.344$ & $0.2 \%$ & $82.6 \%$ & $17.2 \%$ \\
& BFS & $120.4\pm7.1$ & $99.2$  & $100.9$ & $0.359\pm0.047$ & $0.244$ & $0.1 \%$ & $77.5 \%$ & $22.4 \%$ \\
& FFS & $10.6\pm4.2$ & $35.5$  & $30.0$ & $0.000\pm0.000$ & $0.000$ & $0.0 \%$ & $0.0 \%$ & $100.0 \%$ \\
& EXS & $131.2\pm6.0$ & $91.4$  & $95.4$ & $0.388\pm0.019$ & $0.284$ & $0.2 \%$ & $82.0 \%$ & $17.8 \%$ \\
\noalign{\smallskip}\hline
\end{tabular}
\end{table}

\begin{figure}[!t]
\centering
\subfigure[\textit{Collab} network]{\includegraphics[width=0.45\columnwidth]{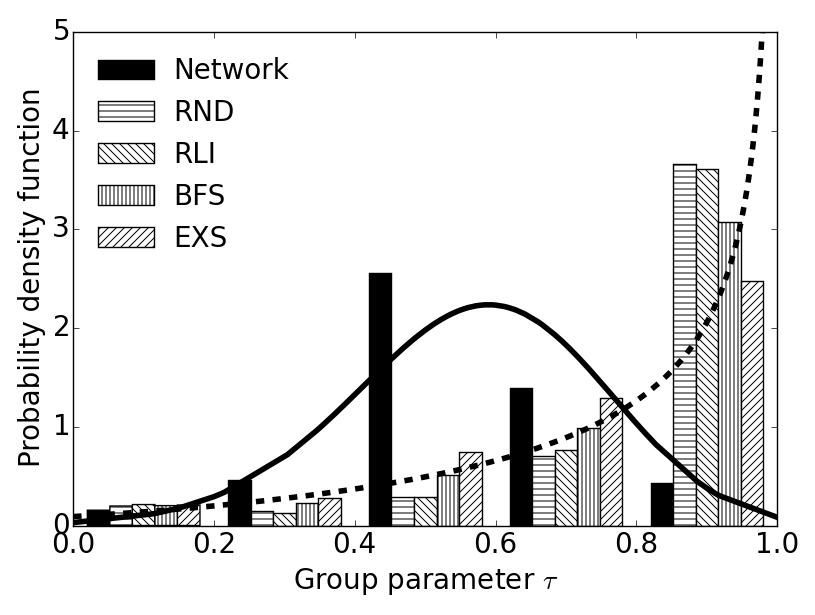}\label{subfig:tca}} \quad
\subfigure[\textit{PGP} network]{\includegraphics[width=0.45\columnwidth]{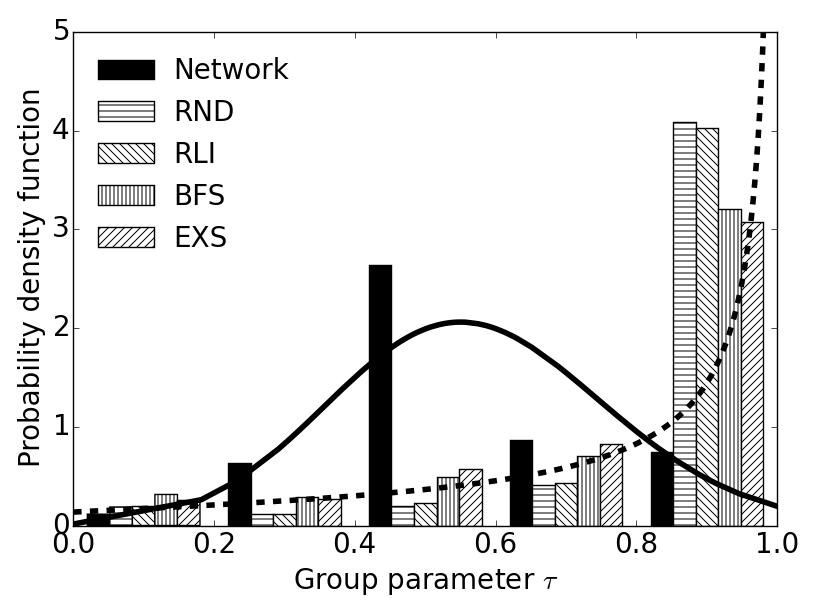}\label{subfig:tpgp}} \\
\subfigure[\textit{P2P} network]{\includegraphics[width=0.45\columnwidth]{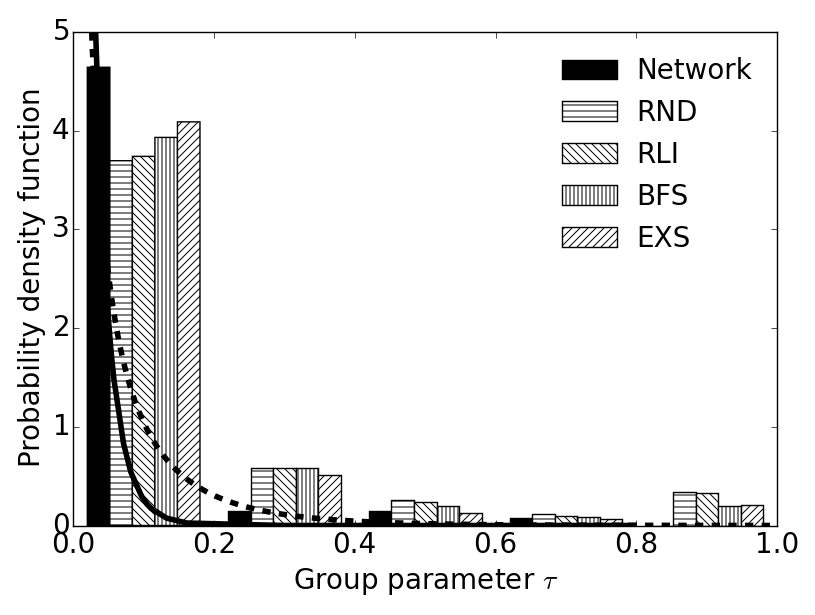}\label{subfig:tp2p}} \quad
\subfigure[\textit{Citation} network]{\includegraphics[width=0.45\columnwidth]{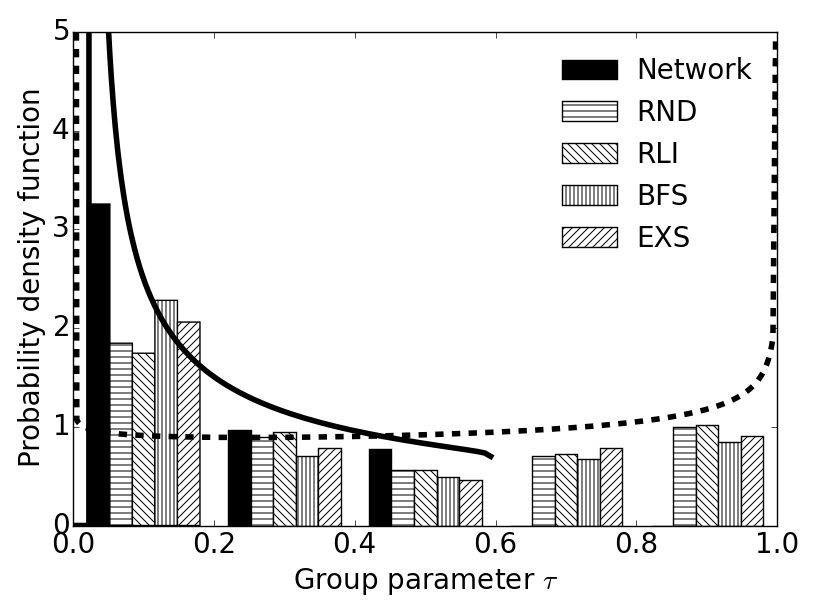}\label{subfig:tcit}}
\caption{Distributions of group parameter $\tau$ for the original networks and their sampled representatives obtained with selected sampling techniques (see text). Histograms are derived by standard equidistant binning, while the estimates of a beta distribution for the original (solid lines) and sampled networks (dashed lines) are merely a guide for the eye. Notice that sampled networks are characterized by denser groups with notably larger $\tau$ than the original ones. Groups are more community-like in the case of social networks (see \subref{subfig:tca} and \subref{subfig:tpgp}), while less module-like in the case of information networks (see \subref{subfig:tp2p} and~\subref{subfig:tcit}).
}
\label{fig:tau-hists}
\end{figure}

\begin{figure}[!t]
\centering
\subfigure[Number of groups \#]{\includegraphics[width=0.45\columnwidth]{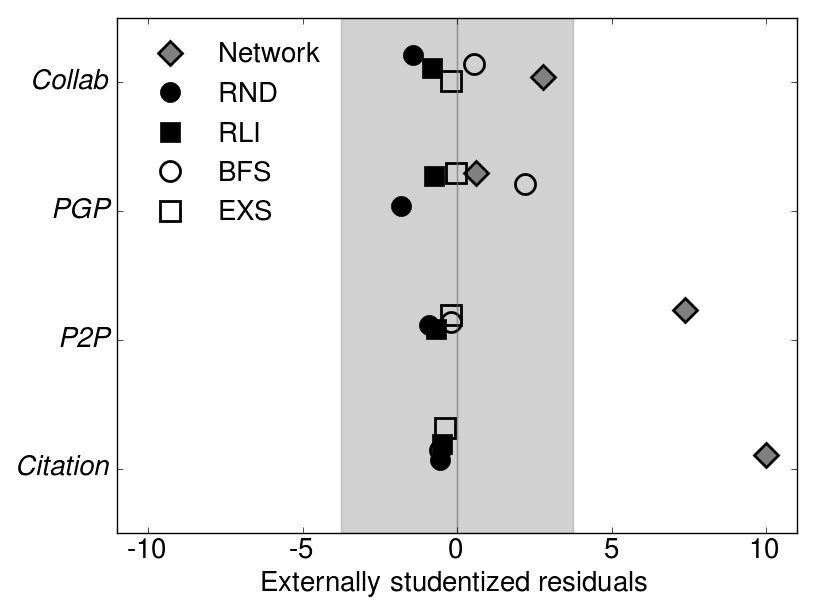}\label{subfig:nro}} \quad
\subfigure[Group parameter $\left\langle \tau \right\rangle$]{\includegraphics[width=0.45\columnwidth]{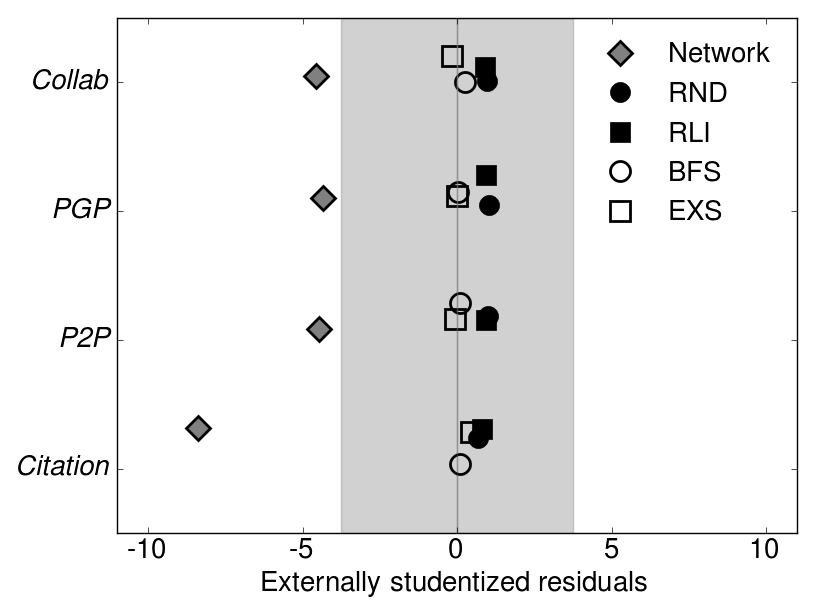}\label{subfig:tauo}} \\
\caption{Statistical comparison of \subref{subfig:nro} number of groups and \subref{subfig:tauo} mean group parameter $\tau$ for the original networks and their sampled representatives obtained with selected sampling techniques (see text). We show externally studentized residuals that measure the consistency of each network with the rest and expose statistically significant inconsistencies between the networks with two-tailed Student $t$-test at $P\mbox{-value}$ of $0.1$ (shaded regions correspond to $90\%$ confidence intervals). Notice that original networks reveal more groups (see \subref{subfig:nro}) with significantly smaller parameter $\tau$ (see \subref{subfig:tauo}) than the sampled networks.
}
\label{fig:stauo}
\end{figure}

Indeed, the majority of groups found in sampled social networks are community-like, which reflects in the parameter $\tau > 0.7$. In sampled information networks the number of mixtures decreases and communities appear, thus $\tau$ is larger than in the original networks. Fig.~\ref{fig:tau-hists} shows a clear difference in the distribution of $\tau$ between the original and sampled networks. Furthermore, to confirm that differences exist between the structure of the original and sampled networks, we compute externally studentized residuals, where we include the value of considered property of the original network in computing the mean over different sampling techniques. We compare the number of groups and the parameter $\left<\tau\right>$ for the original networks and their samples (Fig.~\ref{fig:stauo}). The results prove that the original networks contain a significantly larger number of groups with significantly smaller $\left<\tau\right>$ than the sampled networks. Yet, larger parameter $\tau$ and consequently more community-like groups in sampled social networks and less module-like groups in sampled information networks indicate clear changes in the network structure introduced by sampling. We conclude that these changes occur regardless of the network type or the adopted sampling technique.

Notice that the largest $\tau$ and thus the strongest characterization by community-like groups is revealed in the sampled networks obtained with both random selection techniques, RND and RLI. In RND nodes with higher degrees are more likely to be selected to the sample by definition, while RLI is biased in a similar way~\cite{ANK11}. Thus, densely connected groups of nodes have a higher chance of being included in the sampled network, while sparse parts of the networks remain unsampled. On the other hand, BFS and EXS sample the broad neighborhood of a randomly selected seed node and thus the sampled network represents a connected component. In the case of BFS, all nodes and links of some particular part of the original network are sampled. The latter is believed to be representative of the entire network~\cite{KMT10}, yet BFS is biased towards sampling nodes with higher degree~\cite{NW01} and overestimates the clustering coefficient, especially in information networks~\cite{LKJ06}. On the other hand, EXS ensures the smallest partition distance among several other sampling techniques, which means that nodes grouped together in communities of sampled network are also in the same community in the original network~\cite{MBT10}. Therefore, the stronger characterization by community-like groups in sampled networks can also be explained by the definition and behavior of the sampling techniques.

\subsection{\label{subsec:large}Group structure of a large network}
Due to the relatively high time complexity of the node group extraction framework, we consider only networks with a few thousand nodes. However, our previous study~\cite{BSB14} proved that the size of the original network does not affect the accuracy of the sampling. Still, for a complete analysis, we also inspect the changes in node group structure introduced by sampling of a large \textit{NotreDame} network with more than a million links. Due to the simplicity and execution time, we present the analysis for two sampling techniques, RND from random selection and BFS from network exploration category. We also limit the number of groups extracted from the networks to $100$ (i.e., we consider top $100$ most significant groups with respect to the objective function).

The \textit{NotreDame} data are collected from the web pages of the University of Notre Dame -- \textit{nd.edu} domain in $1999$. The network contains $325$,$729$ nodes representing individual web pages, while $1$,$497$,$134$ links denote hyperlinks among them.

Table~\ref{tbl:wnd} shows the properties of groups, found in the original and sampled networks. The samples consist of smaller groups, still the mean size $s$ remains larger than the mean size $t$. The majority of groups extracted from the original network are module-like, which reflects in the parameter $\tau$ slightly larger than $0$. On the other hand, the changes introduced by sampling are clear, since the samples contain less modules, which is revealed by a larger parameter $\tau$. These findings are consistent with the results on smaller networks from previous sections. The \textit{NotreDame} as an information network expectedly consists of densely linked groups similar to modules, while the structure of sampled networks exhibits stronger characterization by community-like groups. That is again irrespective of the adopted sampling technique.

\begin{table}[!t]
\scriptsize
\centering
\caption{Groups of nodes extracted from the original \textit{NotreDame} network and its sampled representatives over $100$ realizations of selected sampling techniques (see text). We report the number of groups \#, the mean group size $s$, the mean pattern size $t$, the mean group parameter $\tau$ and standard deviation, the median group parameter denoted $m_{\tau}$ and the distribution over different types of groups. Notice that sampled networks expectedly consist of smaller groups, but with larger $\tau$ than original network (see $\left\langle \tau \right\rangle$ and $m_{\tau}$).}
\label{tbl:wnd}
\begin{tabular}{crrrrrrrr}
\hline\noalign{\smallskip}
\multicolumn{1}{c}{Sampling} & \multicolumn{5}{c}{Group} & \multicolumn{1}{c}{Community} & \multicolumn{1}{c}{Mixture} & \multicolumn{1}{c}{Module}\\
& \multicolumn{1}{c}{\#} & \multicolumn{1}{c}{$\left\langle s \right\rangle$} & \multicolumn{1}{c}{$\left\langle t \right\rangle$} & \multicolumn{1}{c}{$\left\langle \tau \right\rangle$} & \multicolumn{1}{c}{$m_{\tau}$}& \multicolumn{3}{c}{Distribution \%} \\
\noalign{\smallskip}\hline\noalign{\smallskip}
/ & $100$ & $876.8$  & $403.6$ & \multicolumn{1}{l}{$0.030$} & $0.028$ & $0.0 \%$ & $99.0 \%$ & $1.0 \%$ \\
\noalign{\smallskip}\hline\noalign{\smallskip}
RND & $100$ & $302.5$  & $271.7$ & $0.369\pm0.010$ & $0.364$ & $0.0 \%$ & $ 100.0\%$ & $ 0.0\%$ \\
BFS & $100$ & $411.6$  & $251.7$ & $0.135\pm0.030$ & $0.119$ & $0.0 \%$ & $ 99.5\%$ & $ 0.5\%$ \\
\noalign{\smallskip}\hline
\end{tabular}
\end{table}

\section{\label{sec:conclusion}Conclusion}
In this paper, we study the presence of characteristic groups of nodes like communities and modules in different social and information networks. We observe the groups of the original networks and analyze the changes in the group structure introduced by the network sampling.

The results first reveal noticeable differences in the group structure of original social and information networks. Nodes in social networks form smaller community-like groups, while information networks are better characterized by larger modules. After applying network sampling techniques, sampled networks expectedly contain fewer and smaller groups. However, the sampled networks exhibit stronger characterization by community-like groups than the original networks. We have shown that the changes in the node group structure introduced by sampling occur regardless of the network type and consistently across different sampling techniques. Since networks commonly considered in the literature are inevitably just a sampled representative of its real-world analogue, some results, such as rich community structure found in these networks, may be influenced by or are merely an artifact of sampling.

Our future work will mainly focus on larger real-world networks, including other types of networks like biological and technological. Moreover, we will further analyze the changes in the node group structure introduced by sampling and explore techniques that could overcome observed deficiencies.

\section*{Acknowledgment}
This work has been supported in part by the Slovenian Research Agency \textit{ARRS} within the Research Program No. P2-0359, by the Slovenian Ministry of Education, Science and Sport Grant No. 430-168/2013/91, and by the European Union, European Social Fund.

\bibliographystyle{elsarticle-num}

\begin{thebibliography}{50}
\expandafter\ifx\csname url\endcsname\relax
  \def\url#1{\texttt{#1}}\fi
\expandafter\ifx\csname urlprefix\endcsname\relax\def\urlprefix{URL }\fi
\expandafter\ifx\csname href\endcsname\relax
  \def\href#1#2{#2} \def\path#1{#1}\fi

\bibitem{LKJ06}
S.~H. Lee, P.~J. Kim, H.~Jeong, Statistical properties of sampled networks,
  Phys. Rev. E 73~(1) (2006) 016102.

\bibitem{BA99}
A.~L. Barab{\'a}si, R.~Albert, Emergence of scaling in random networks, Science
  286~(5439) (1999) 509--512.

\bibitem{Fre77}
L.~Freeman, A set of measures of centrality based on betweenness, Sociometry
  40~(1) (1977) 35--41.

\bibitem{Newman02}
M.~E.~J. Newman, Assortative mixing in networks, Phys. Rev. Lett. 89~(20)
  (2002) 208701.

\bibitem{SWM05}
M.~P.~H. Stumpf, C.~Wiuf, R.~M. May, Subnets of scale-free networks are not
  scale-free: sampling properties of networks, P. Natl. Acad. Sci. USA 102~(12)
  (2005) 4221--4224.

\bibitem{SCBFGP12}
S.-W. Son, C.~Christensen, G.~Bizhani, D.~V. Foster, P.~Grassberger,
  M.~Paczuski, Sampling properties of directed networks, Phys. Rev. E 86~(4)
  (2012) 046104.

\bibitem{LF06}
J.~Leskovec, C.~Faloutsos, Sampling from large graphs, in: Proceedings of the
  12th ACM SIGKDD international conference on Knowledge discovery and data
  mining, ACM, 2006, pp. 631--636.

\bibitem{LKF05}
J.~Leskovec, J.~Kleinberg, C.~Faloutsos, Graphs over time: Densification laws,
  shrinking diameters and possible explanations, in: Proceedings of the 11th
  ACM SIGKDD International Conference on Knowledge Discovery and Data Mining,
  ACM, 2005, pp. 177--187.

\bibitem{WS98}
D.~J. Watts, S.~H. Strogatz, Collective dynamics of ‘small-world’networks,
  Nature 393~(6684) (1998) 440--442.

\bibitem{ANK11}
N.~Ahmed, J.~Neville, R.~R. Kompella, Network sampling via edge-based node
  selection with graph induction, Tech. rep., Purdue University (2011).

\bibitem{PM13}
H.~Park, S.~Moon, Sampling bias in user attribute estimation of osns, in:
  Proceedings of the 22nd international conference on World Wide Web companion,
  International World Wide Web Conferences Steering Committee, 2013, pp.
  183--184.

\bibitem{LBCX03}
A.~Lakhina, J.~W. Byers, M.~Crovella, P.~Xie, Sampling biases in ip topology
  measurements, in: Proceedings of the 22nd Annual Joint Conference of the IEEE
  Computer and Communications, Vol.~1, IEEE, 2003, pp. 332--341.

\bibitem{MB11}
A.~S. Maiya, T.~Y. Berger-Wolf, Benefits of bias: Towards better
  characterization of network sampling, in: Proceedings of the 17th ACM SIGKDD
  international conference on Knowledge discovery and data mining, ACM, 2011,
  pp. 105--113.

\bibitem{LWZY07}
B.~Long, X.~Wu, Z.~Zhang, P.~S. Yu, Community learning by graph approximation,
  in: Proceedings of 7th IEEE International Conference on Data Mining, IEEE,
  2007, pp. 232--241.

\bibitem{GN02}
M.~Girvan, M.~E.~J. Newman, Community structure in social and biological
  networks, P. Natl. Acad. Sci. USA 99~(12) (2002) 7821--7826.

\bibitem{SC11}
J.~Scott, P.~J. Carrington, The SAGE handbook of social network analysis, SAGE
  publications, 2011.

\bibitem{RB08}
M.~Rosvall, C.~T. Bergstrom, Maps of random walks on complex networks reveal
  community structure, P. Natl. Acad. Sci. {USA} 105~(4) (2008) 1118--1123.

\bibitem{ADP06}
A.~Arenas, A.~D{\'\i}az-Guilera, C.~J. P{\'e}rez-Vicente, Synchronization
  reveals topological scales in complex networks, Phys. Rev. Lett. 96~(11)
  (2006) 114102.

\bibitem{WH04}
F.~Wu, B.~A. Huberman, Finding communities in linear time: a physics approach,
  Eur. Phys. J. B 38~(2) (2004) 331--338.

\bibitem{RB07}
M.~Rosvall, C.~T. Bergstrom, An information-theoretic framework for resolving
  community structure in complex networks, P. Natl. Acad. Sci. {USA} 104~(18)
  (2007) 7327--7331.

\bibitem{RAK07}
U.~N. Raghavan, R.~Albert, S.~Kumara, Near linear time algorithm to detect
  community structures in large-scale networks, Phys. Rev. E 76~(3) (2007)
  036106.

\bibitem{SB11}
L.~{\v{S}}ubelj, M.~Bajec, Unfolding communities in large complex networks:
  Combining defensive and offensive label propagation for core extraction,
  Phys. Rev. E 83~(3) (2011) 036103.

\bibitem{Fortunato10}
S.~Fortunato, Community detection in graphs, Phys. Rep. 486~(3) (2010) 75--174.

\bibitem{RCCLP04}
F.~Radicchi, C.~Castellano, F.~Cecconi, V.~Loreto, D.~Parisi, Defining and
  identifying communities in networks, P. Natl. Acad. Sci. {USA} 101~(9) (2004)
  2658--2663.

\bibitem{NL07}
M.~E.~J. Newman, E.~A. Leicht, Mixture models and exploratory analysis in
  networks, P. Natl. Acad. Sci. {USA} 104~(23) (2007) 9564.

\bibitem{PSR10}
S.~Pinkert, J.~Schultz, J.~Reichardt, Protein interaction networks—more than
  mere modules, PLoS Computational Biology 6~(1) (2010) e1000659.

\bibitem{SB12}
L.~{\v{S}}ubelj, M.~Bajec, Ubiquitousness of link-density and link-pattern
  communities in real-world networks, Eur. Phys. J. B 85~(1) (2012) 1--11.

\bibitem{RW07}
J.~Reichardt, D.~R. White, Role models for complex networks, Eur. Phys. J. B
  60~(2) (2007) 217--224.

\bibitem{SBB13}
L.~{\v{S}}ubelj, N.~Blagus, M.~Bajec, Group extraction for real-world networks:
  The case of communities, modules, and hubs and spokes, in: Proceedings of the
  International Conference on Network Science, 2013, pp. 152--153.

\bibitem{SRR12}
M.~Salehi, H.~R. Rabiee, A.~Rajabi, Sampling from complex networks with high
  community structures, Chaos 22~(2) (2012) 023126.

\bibitem{MBT10}
A.~S. Maiya, T.~Y. Berger-Wolf, Sampling community structure, in: Proceedings
  of the 19th international conference on World wide web, ACM, 2010, pp.
  701--710.

\bibitem{LLPP14}
B.~Lu{\v{z}}ar, Z.~Levnaji{\'c}, J.~Povh, M.~Perc, Community structure and the
  evolution of interdisciplinarity in slovenia's scientific collaboration
  network, PloS One 9~(4) (2014) e94429.

\bibitem{perc14}
M.~Perc, The matthew effect in empirical data, J. Roy. Soc. Interface 11~(98)
  (2014) 20140378.

\bibitem{BP98}
S.~Brin, L.~Page, The anatomy of a large-scale hypertextual web search engine,
  Comput. Networks ISDN Syst. 30~(1) (1998) 107--117.

\bibitem{BSB14}
N.~Blagus, L.~{\v{S}}ubelj, M.~Bajec, Assessing the effectiveness of real-world
  network simplification, Physica A 413 (2014) 134--146.

\bibitem{ANK12}
N.~K. Ahmed, J.~Neville, R.~Kompella, Network sampling: from static to
  streaming graphs, e-print {arXiv:11211.3412}.

\bibitem{KMT10}
M.~Kurant, A.~Markopoulou, P.~Thiran, On the bias of {BFS}, in: Proceedings of
  the 22nd International Teletraffic Congress, IEEE, 2010, pp. 1--8.

\bibitem{goodman61}
L.~A. Goodman, Snowball sampling, Ann. Math. Stat. (1961) 148--170.

\bibitem{ZLZ11}
Y.~Zhao, E.~Levina, J.~Zhu, Community extraction for social networks, P. Natl.
  Acad. Sci. 108~(18) (2011) 7321--7326.

\bibitem{Weiss}
G.~Weiss, L.~{\v{S}}ubelj, nets-nodegroups v1.0,
  \url{http://dx.doi.org/10.5281/zenodo.11589} (2014).
\newblock \href {http://dx.doi.org/10.5281/zenodo.11589}
  {\path{doi:10.5281/zenodo.11589}}.

\bibitem{SZBB14}
L.~{\v{S}}ubelj, S.~{\v{Z}}itnik, N.~Blagus, M.~Bajec, Node mixing and group
  structure of complex software neworks, Advs. Complex Syst. 17 (2014) 1450022.

\bibitem{jaccard1901}
P.~Jaccard, \'{E}tude comparative de la distribution florale dans une portion
  des alpes et du jura, Bull. Soc. Vaud. Sci. Nat. 37 (1901) 547--579.

\bibitem{RN03}
S.~Russel, P.~Norvig, Artificial Intelligence: A Modern Approach (second
  edition), Upper Saddle River, N. J.: Prentice Hall, 2003.

\bibitem{ER59}
P.~Erd{\H{o}}s, A.~R{\'e}nyi, On random graphs i., Publ. Math. Debrecen 6
  (1959) 290--297.

\bibitem{PDFV05}
G.~Palla, I.~Der{\'e}nyi, I.~Farkas, T.~Vicsek, Uncovering the overlapping
  community structure of complex networks in nature and society, Nature
  435~(7043) (2005) 814--818.

\bibitem{RSMOB02}
E.~Ravasz, A.~L. Somera, D.~A. Mongru, Z.~N. Oltvai, A.-L. Barab{\'a}si,
  Hierarchical organization of modularity in metabolic networks, Science
  297~(5586) (2002) 1551--1555.

\bibitem{BPDA04}
M.~Bogu{\~n}{\'a}, R.~Pastor-Satorras, A.~D{\'\i}az-Guilera, A.~Arenas, Models
  of social networks based on social distance attachment, Phys. Rev. E 70~(5)
  (2004) 056122.

\bibitem{NP03}
M.~E.~J. Newman, J.~Park, Why social networks are different from other types of
  networks, Phys. Rev. E 68~(3) (2003) 036122.

\bibitem{SFB14}
L.~{\v{S}}ubelj, D.~Fiala, M.~Bajec, Network-based statistical comparison of
  citation topology of bibliographic databases, Sci. Rep. 4 (2014) 6496.

\bibitem{CW82}
R.~D. Cook, S.~Weisberg, Residuals and influence in regression, Mg. Stat. Pro.

\bibitem{NW01}
M.~Najork, J.~L. Wiener, Breadth-first crawling yields high-quality pages, in:
  Proceedings of the 10th international conference on World Wide Web, ACM,
  2001, pp. 114--118.

\end{thebibliography}

\end{document}